# Femtosecond Engineering of PolyMethylPentene (PMP) Nonlinearity via NA, Spin/orbital angular momentum (OAM), and In-Situ SC Diagnostics

Siyang Zheng[1], Guangyu Zhu[2, *], Walter Perrie[3], Dongzhi Wang[5], Runzhen Zhou[2], Juan Ignacio Ahuir-Torre[4] and Puxiang Lai[1]

[1] *Department of Biomedical Engineering, Hong Kong Polytechnic University, HK, CN*
[2] *School of Electrical Engineering, Chongqing University of Arts and Sciences, CQ, 402160, CN*
[3] *Laser Group, School of Engineering, Brownlow Street, University of Liverpool, L69 3GQ, UK*
[4] *General Engineering Research Institute, Faculty of Engineering and Technology, John Moores University, Liverpool, L3 3AF, UK*
[5] *TPM3D Direct Manufacturing Co., Ltd., Yancheng, Jiangsu, 224000, CN*
**zchuguangyu@cqwu.edu.cn.*

**Abstract:** Despite the growing interest in femtosecond vortex beams for photonic device fabrication, the role of orbital angular momentum (OAM) in mediating higher-order nonlinear interactions in transparent polymers remains poorly understood, particularly regarding the tensorial nature of the third-order susceptibility and its influence on energy deposition and structural symmetry. To address this gap, this work investigates the interaction of femtosecond vortex beams with Polymethylpentene (PMP) polymers, establishing OAM as a controllable parameter for precision beam-shaping and energy deposition in ultrafast laser micromachining and photonic device fabrication. A 775 nm femtosecond laser) is frequency-doubled to 387.5 nm and phase-shaped by a reflective liquid crystal-spatial light modulator, with OAM imprinted via computer-generated holograms and independently controlled spin via a quarter-wave plate. Theoretical modeling of the tensorial nonlinear polarization, combined with systematic experiments on low-NA filamentation, supercontinuum spectroscopy using an advanced spectrometer, and high-NA inscription with NA = 0.4 and 0.7, reveals that the same $\chi^{(3)}$ tensor components govern all regimes.The transition from perturbative to dissipative behavior is determined solely by whether the local intensity exceeds the plasma formation threshold. Through an analytical framework based on the isotropic material model, the key innovation lies in identifying that $\chi_{1122}$ governs the energy deposition and plasma threshold, whereas $\chi_{1221}$ mediates the transverse nonlinear current and helicity-dependent asymmetry, thereby converting phase topology into permanent chiral structures. This enables OAM-controlled structuring with sub-micrometer feature sizes (~0.43 µm diameter, ~200 nm² area) and localized intensities reaching $10^{14}$–$10^{15}$ W/cm², providing an engineering framework for deterministic fabrication of chiral waveguides and photonic devices through precise control of topological charge and focusing conditions.

## References and links

---

## 1. Introduction

In recent years, femtosecond vortex beams have attracted considerable research interest owing to their helical phase fronts and well-defined orbital angular momentum (OAM), which introduce new degrees of freedom in ultrafast light–matter interaction and underpin applications ranging from super-resolution imaging and optical manipulation to high-capacity optical communications [1-5]. Significant progress has been made in the characterization and recognition of structured beams [6], including demonstrations of compact vortex sources [7], and integrated waveguide-based generators fabricated via femtosecond laser direct writing [8-10], which have greatly enhanced the practical accessibility and integrability of these beams. In parallel, laser material processing has emerged as a particularly active frontier, where femtosecond vortex beams have been exploited for multiphoton polymerization and internal structuring of transparent dielectrics [11–13], uniform multi-filament array formation [14], and fabrication of complex three-dimensional chiral microstructures [15], leveraging their unique non-Gaussian energy distributions and phase-dependent nonlinear coupling [16–18]. Despite these advances, the role of OAM in higher-order nonlinear interactions within dielectrics remains insufficiently understood, particularly regarding energy redistribution, helicity dynamics, and the $\chi^{(3)}$-mediated transition from perturbative to consumptive regimes.

Among bulk substrates, undoped polymers offer distinct advantages for femtosecond inscription, including low cost, mechanical flexibility, and broad spectral transparency.

Polymethylpentene (PMP), which exhibits transparency from the ultraviolet to the far-infrared region [19], has recently emerged as a promising candidate, demonstrating diffraction efficiencies an order of magnitude higher than those of Poly(methyl methacrylate) (PMMA) in volume Bragg grating inscription—an attribute ascribed to its large free volume promoting localized contraction with $\Delta n \approx 5\times10^{-4}$ under low-NA focusing, while NA = 0.4 inscription achieves $\Delta n \approx 0.01$ in thick gratings, implicating two-photon absorption as the dominant mechanism [20–22]. Nevertheless, the third-order susceptibility $\chi^{(3)}$ [23], which governs self-focusing [24], plasma generation [25], and the perturbative-to-consumptive transition, remains poorly characterized in PMP under vortex beam geometries. Furthermore, existing studies have treated filamentation [26], supercontinuum generation [27], and inscription [28] separately, lacking a unified framework connecting these processes through the tensor nature of $\chi^{(3)}$ ) in transparent media. In particular, for isotropic materials, the distinct roles of individual tensor components $\chi_{1122}$ and $\chi_{1221}$ in mediating energy deposition and structural modification remain less clearance. To address these gaps, the present study systematically investigates OAM-dependent $\chi^{(3)}$-mediated nonlinear interactions and their implications for the energy redistribution and the structural transformation in transparent polymers.

This work addresses the gap through combined theory and experiment, establishing OAM as a structuring parameter for transparent thermoplastics, where filamentation, supercontinuum, and inscription share a common $\chi^{(3)}$origin linking phase topology to permanent structures. The findings hold significance for laser engineering sector, offering topological control over energy deposition across regimes via tensorial nonlinear response. In addition, here we employ $m_{2\omega}$ for the topological charge after the extracavity beta-barium borate (BBO) doubling stage; in ordinary discussion, the topological parameter is denoted simply as m with no subscript.

## 2. Experimental details

The experimental layout is shown in Fig. 1. A 775 nm femtosecond laser (Clark-MXR-CPA 2010) was attenuated, expanded, and phase-shaped by a reflective liquid-crystal spatial light modulator (SLM) (Hamamatsu X-10468-02) for high-order tailored wavefront control, then frequency-doubled to 387.5 nm in a BBO crystal (~10% efficiency). A 4f system relayed the SLM plane to the objective pupil, with a spatial filter at the Fourier plane removing higher-order diffraction artifacts [29]. Residual 775 nm light was blocked by dielectric mirrors, while a pick-off directed a small portion to a beam profiler (Spiricon SP620U) for in situ focal-spot monitoring. Polished PMP samples were mounted on a three-axis air-bearing stage (Aerotech). OAM was imprinted at 775 nm via computer-generated holograms (LabVIEW, see [30]), yielding $2m\hbar$ per photon after doubling, and a quarter-wave plate (±45°) introduced $\pm\hbar$ spin angular momentum. For reflective supercontinuum (SC) spectrum acquisition, the generated continuum was retro-reflected by an aluminised mirror placed directly behind the sample; the collected light passed through the near ultra-violet (NUV) dielectric-coated 45° turning mirror (ahead of the objective), was coupled into a fibre, and delivered to a spectrometer (Andor Shamrock SR-303i, CCD 200–900 nm), enabling in situ spectral feedback from the excitation volume. As summarized in Table 1, a fused silica lens (NA < 0.1), a 50 mm lens, and 0.4 NA and 0.7 NA objectives were used for low-NA work, simulations, inscription, and imaging, respectively.

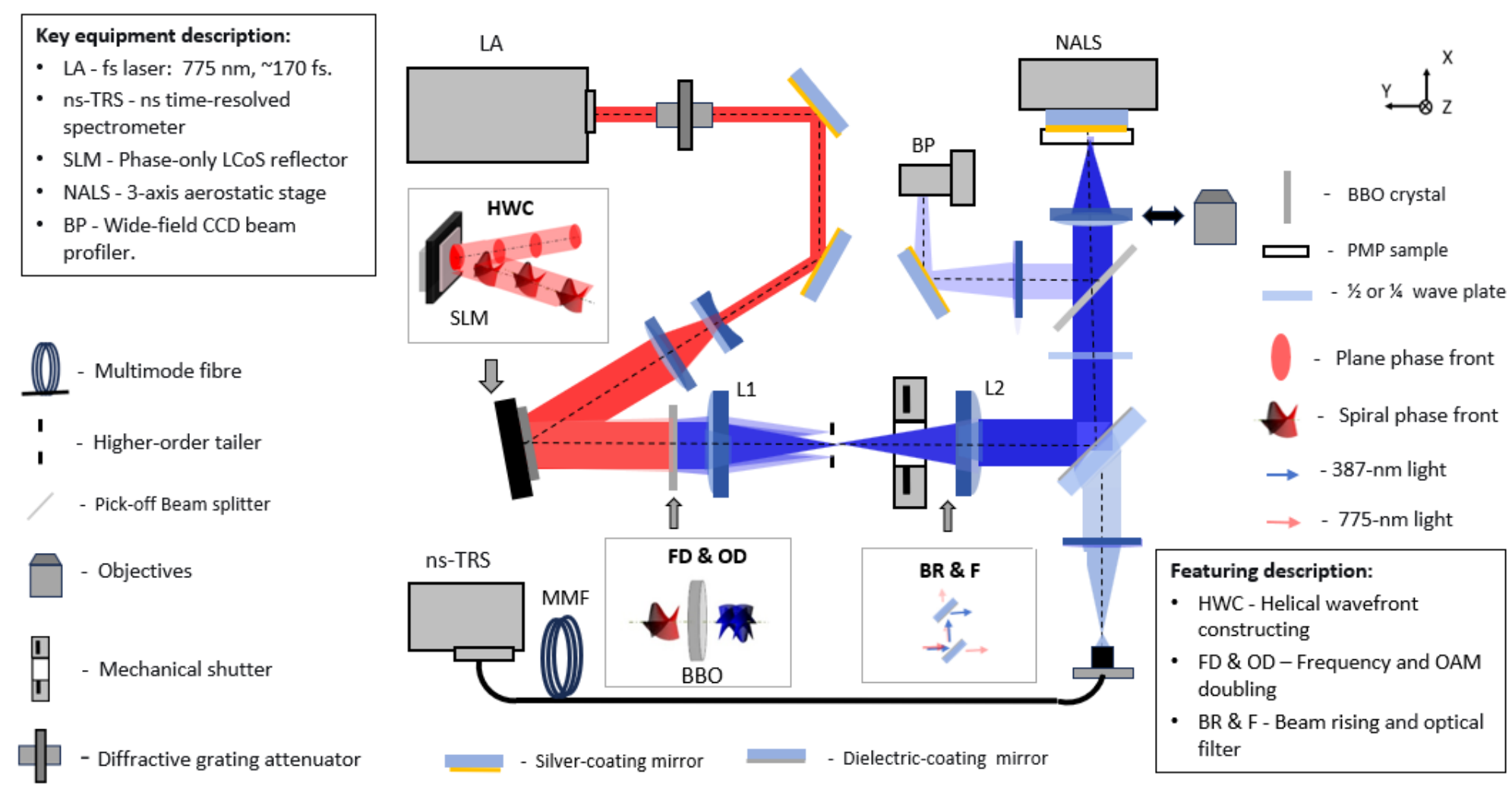

**Fig. 1** Optical setup: OAM is encoded at 775 nm via SLM, doubled to 387.5 nm in BBO, with independent spin control. A spectrometer and beam profiler enable in situ spectral feedback and real-time mode monitoring.

**Table 1.** Summary of focusing optics used in simulations and experiments.

| **Objective/Lens** | **NA** | **Specifications** | **Purpose** |
|---|---|---|---|
| Positive lens | < 0.1 | Fused silica singlet | Low-NA filamentation experiments (Fig.3) and supercontinuum spectroscopy (Fig.4) |
| Positive lens | _ | f = 50 mm | Simulation of vortex beam focusing (170 fs, 1 μJ) to generate annular intensity profiles (Fig.2 and Fig.10) |
| High-NA objective | 0.4 | THORLABS (LMU-20X-NUV) | Moderate-NA inscription for side-view structural characterization (Figs. 5–7) |
| High-NA objective | 0.7 | Mitutoyo M Plan APO 100× | High-NA inscription for near-surface waveguide fabrication and cross-sectional imaging (Figs. 8–9) |

## 3. Theoretical and experimental results with discussion

### 3.1 Analytical Framework for Nonlinear Polarization and Hot Spots in Vortex Beams

A simplified analysis of localized asymmetric intensity enhancement in femtosecond vortex beams is achieved by restricting the study to isotropic media. This reduces mathematical complexity while still capturing the key nonlinear behavior. In such systems, the nonlinear polarization induced by the optical field is expressed in a compact vectorial form, wherein the tensorial nature of the third-order nonlinearity is inherently incorporated. Following the canonical formalism established by Boyd (Equ. (1)) [23], the nonlinear polarization is given by:

$$P = 6\epsilon_0\chi_{1122}(E \cdot E^*)E + 3\epsilon_0\chi_{1221}(E \cdot E^*)E \qquad (1)$$

where $\chi_{1122}$ and $\chi_{1221}$ denote the independent components of the third-order susceptibility which is a fourth-rank tensor that characterize the material anisotropic nonlinear response to the optical field. The tensorial form of Eq. (1) dictates that this nonlinear polarization is not simply proportional to the scalar intensity but depends sensitively on the local polarization and phase topology encoded in the vector field **E**. The incident beam's spatial structure critically determines this nonlinear interaction.

Beams with a helical wavefront, phase factor $e^{im\phi}$ , carry orbital angular momentum (OAM) quantized by topological charge *m*. The central phase singularity yields a Laguerre–Gaussian (LG) ring profile with zero on-axis intensity [31]. In the paraxial regime, the complex electric field amplitude is scalarly described by Equ. (2).

$$E(r,m,\phi) = E_0\left(\frac{1}{\sqrt{m!}}\right)\left(\frac{\sqrt{2}}{\omega_0}r\right)^m e^{(-r/\omega_0 \pm im\phi)} \quad [32] \qquad (2)$$

The corresponding intensity distribution, $I \propto |E|^2$, retains azimuthal symmetry for pure modal states. To illustrate this behavior, the preceding solution is applied to a 170 fs, 1 μJ pulse focused by a 50 mm positive lens. The resulting paraxial intensity profiles are shown in Fig. 2. For the topological charge $m = 1$ , this yields the characteristic annular intensity pattern. Although the time-averaged intensity remains azimuthally symmetric for a pure vortex mode, the underlying vectorial field does not share this symmetry when a linear polarization is imposed across the helical phase front. The local field orientation varies with azimuth, breaking the rotational symmetry of the polarization distribution while leaving the scalar intensity unchanged.

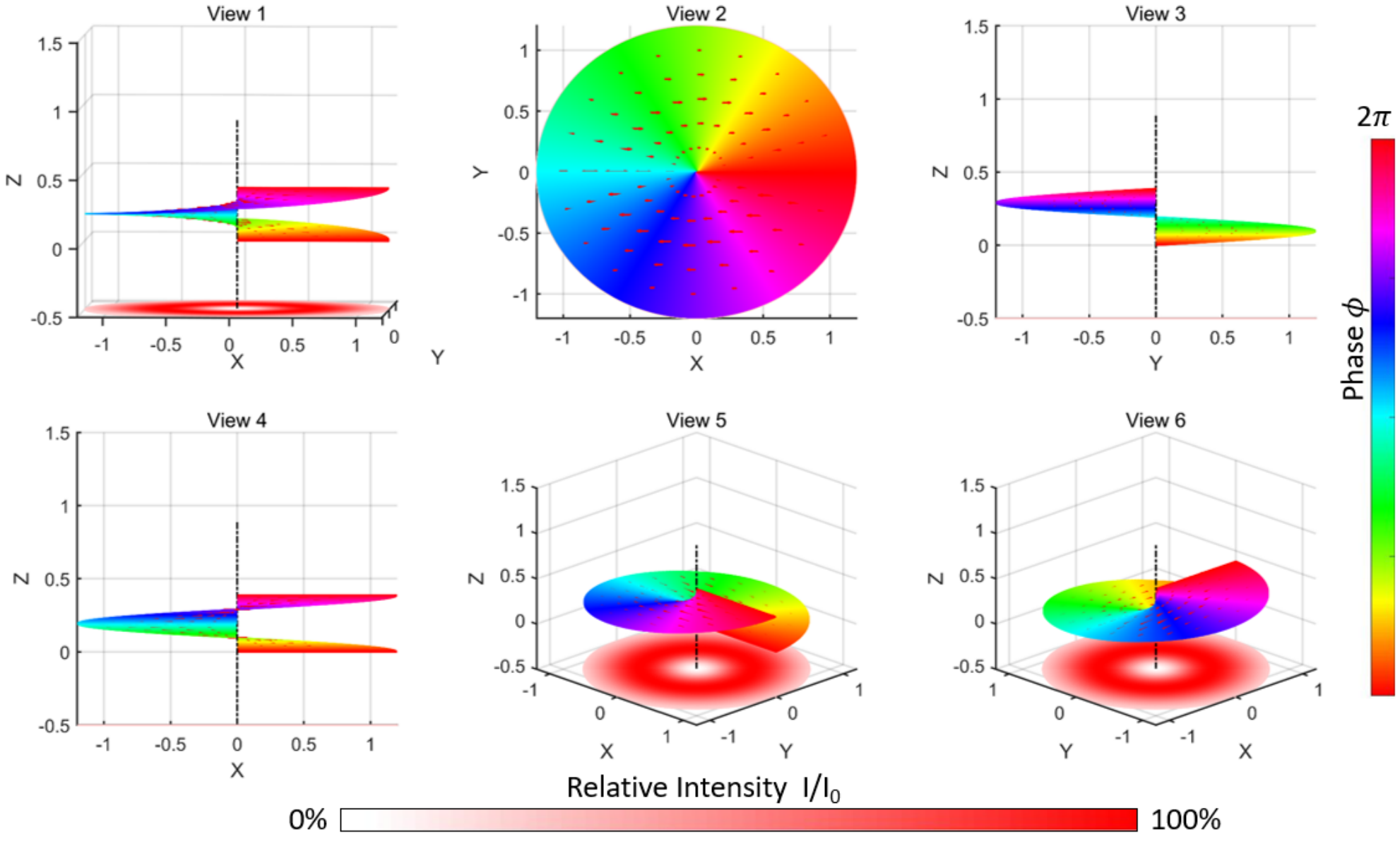

**Fig. 2** Six views of a femtosecond vortex beam (m=1, 170 fs, 1 μJ). Top: helical phase front ($e^{im\phi}$)with azimuthally varying linear polarization orientation (red arrows). Bottom: time-averaged intensity cross-sections showing annular profile with on-axis null. This anisotropic polarization, governed by tensorial susceptibility components ($\chi_{1122}$, $\chi_{1221}$), reshapes $\chi^{(3)}$ via the Kerr effect for nonlinear control.

### 3.2 Low NA inscription with spiral beams carrying both OAM and SAM

To examine whether orbital angular momentum (OAM) and spin angular momentum (SAM) couple in the low-NA filamentation regime, experiments were performed in PMP using beams with topological charges $l = \pm 2$, combined with spin angular momentum $\sigma = \pm 1$ (left- and right-handed circular polarization). If total angular momentum $J = l + \sigma$ governed the interaction, $J$ would be expected to range from −3 to +3 in integer steps. Figure 3 shows optical micrographs of filament cross-sections obtained at $E_p$ = 0.7 μJ for Gaussian and LG beams under linear, right-circular, and left-circular polarizations. Several trends are immediately apparent. For all OAM states, linearly polarized filaments initiate closest to the surface, with onset depth increasing with $l$. Circularly polarized filaments, regardless of handedness, consistently form deeper than their linearly polarized equivalents for the same l, while no difference is observed between right- and left-circular cases.

The stronger coupling observed with linear polarization likely originates from polarization-dependent contributions of the third-order nonlinear susceptibility tensor $\chi^{(3)}$. In isotropic media, the Kerr nonlinearity involves two independent tensor components for linear polarization, but only one under circular polarization [23]. This is consistent with previous reports of polarization-dependent filamentation in PMMA, although those studies were limited to linearly polarized Gaussian beams [33]. For beams carrying OAM, a slight asymmetry appears between $l = +2$ and $l = -2$, yet the addition of SAM ($\sigma = \pm 1$) to a given OAM state produces no measurable effect, as confirmed by repeated measurements. These results show no measurable OAM–SAM coupling at low NA, contradicting the $J = l + \sigma$ rule. Spin–orbit coupling plays little role, while enhanced coupling under linear polarization arises from the tensorial $\chi^{(3)}$ structure, consistent with theoretical study in Sec 3.1.

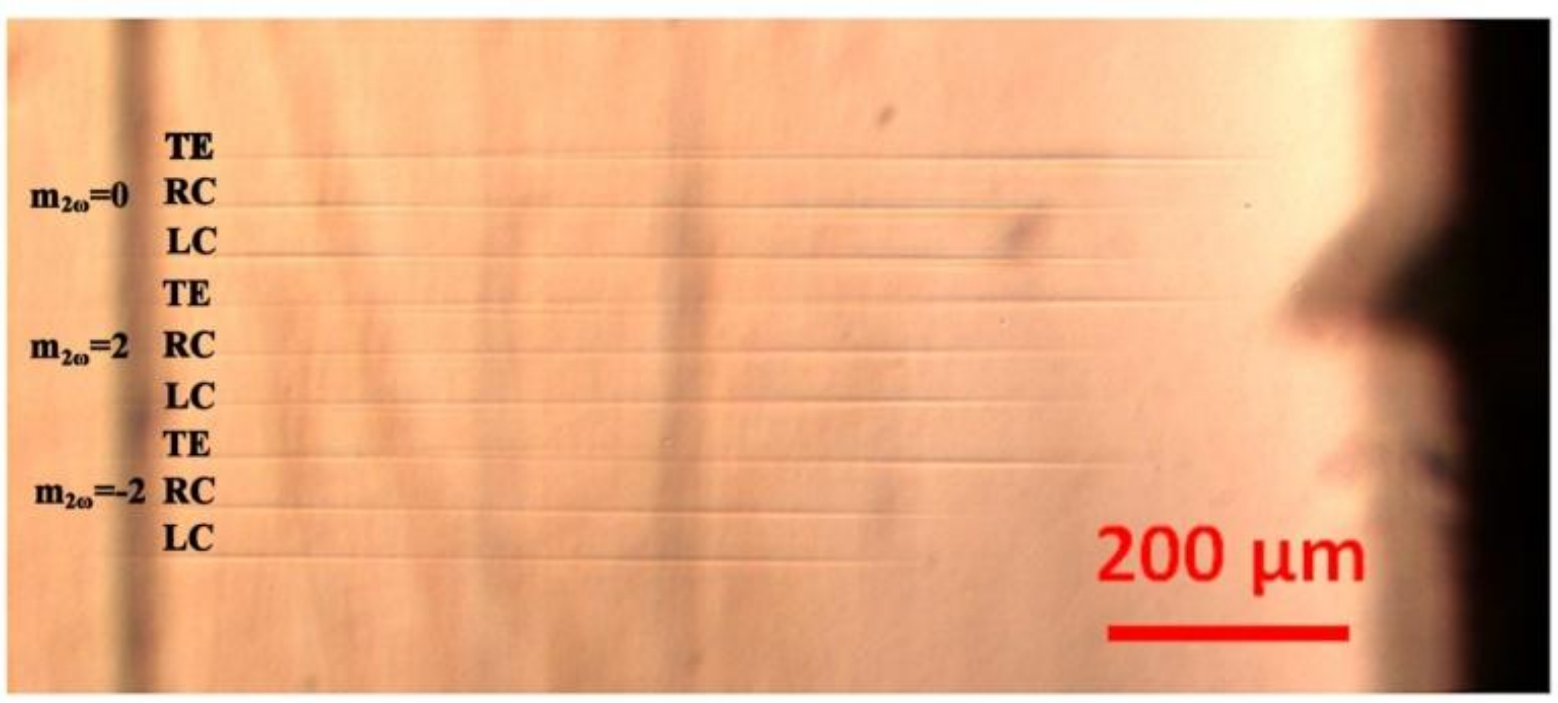

**Fig. 3.** PMP filament cross-sections for Gaussian and LG ($l = \pm 2$) beams under linear, right ($\sigma = +1$), and left ($\sigma = -1$) circular polarizations. Linear polarization initiates closer to the surface; onset depth increases with l, with less OAM–SAM coupling observed.

### 3.3 Supercontinuum (SC) in PMP at 387nm with OAM beams.

Supercontinuum (SC) generation is employed here as a spectroscopic probe of tensorial nonlinear polarization effects, with the incident beam horizontally linearly polarized. In weakly ionized filaments, SC arises from self-phase modulation [34] and self-steepening [35], while two-photon absorption becomes dominant when the photon energy exceeds half the bandgap [36]. Accordingly, two-photon excitation of PMP at 387.5 nm (6.4 eV, well above $Eg/2 \approx 4.3$ eV [37]) produces a blue-enhanced continuum extending from below 300 nm to above 450 nm (Fig. 4(a)).

The central observation concerns the OAM dependence of SC efficiency. As shown in Fig. 4(b), increasing the topological charge $m_{2\omega}$ from 2 to 8 systematically suppresses the SC signal, particularly on the anti-Stokes side (350–370 nm). The integrated intensity over 350–380 nm, plotted against $m_{2\omega}$ in Fig. 4(c), exhibits a linear decay with a slope of −1.88. This linear trend is consistent with the theoretical prediction that critical power scales linearly with topological charge [38]. Although the measured slope is lower than the ideal Laguerre–Gaussian value of 3.46, it remains within the range of previous experimental reports under varying conditions [39–41].

Crucially, the observed linearity, as opposed to higher-order scaling, indicates that the dominant suppression mechanism is the OAM-induced reduction in effective peak intensity, not a change in the order of the nonlinear process itself. This interpretation is further supported by the helical phase gradient of the vortex beam. Through the $\chi_{1221}$ tensor component, this gradient drives a transverse nonlinear current that redistributes energy away from the beam center. The linear decay therefore confirms that the OAM-induced lowering of effective peak intensity is the principal factor responsible for the suppression, although minor corrections from higher-order processes remain possible.

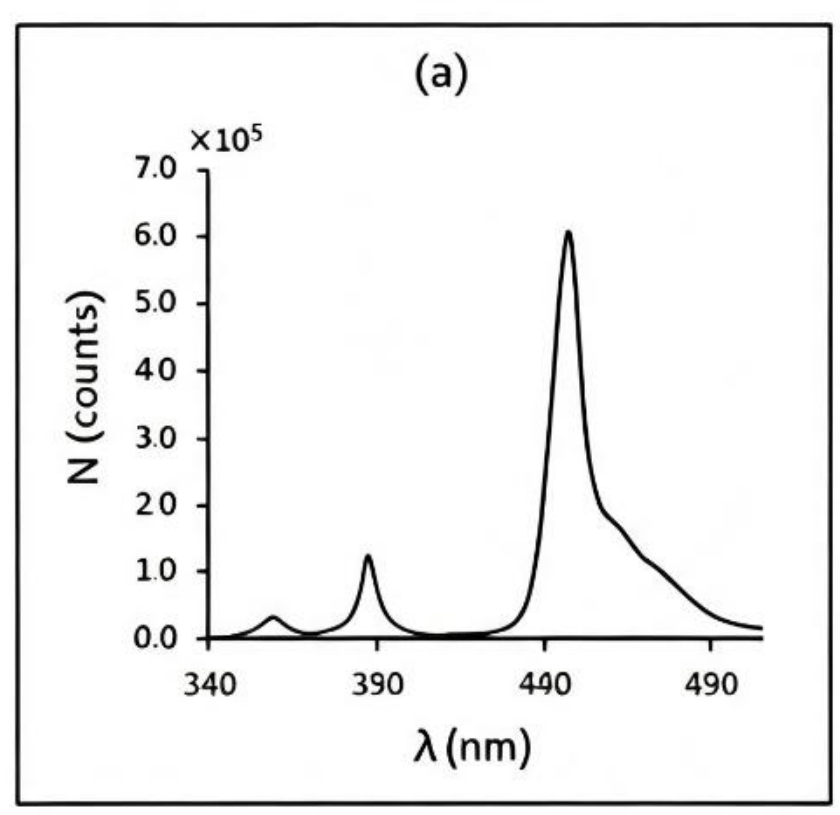

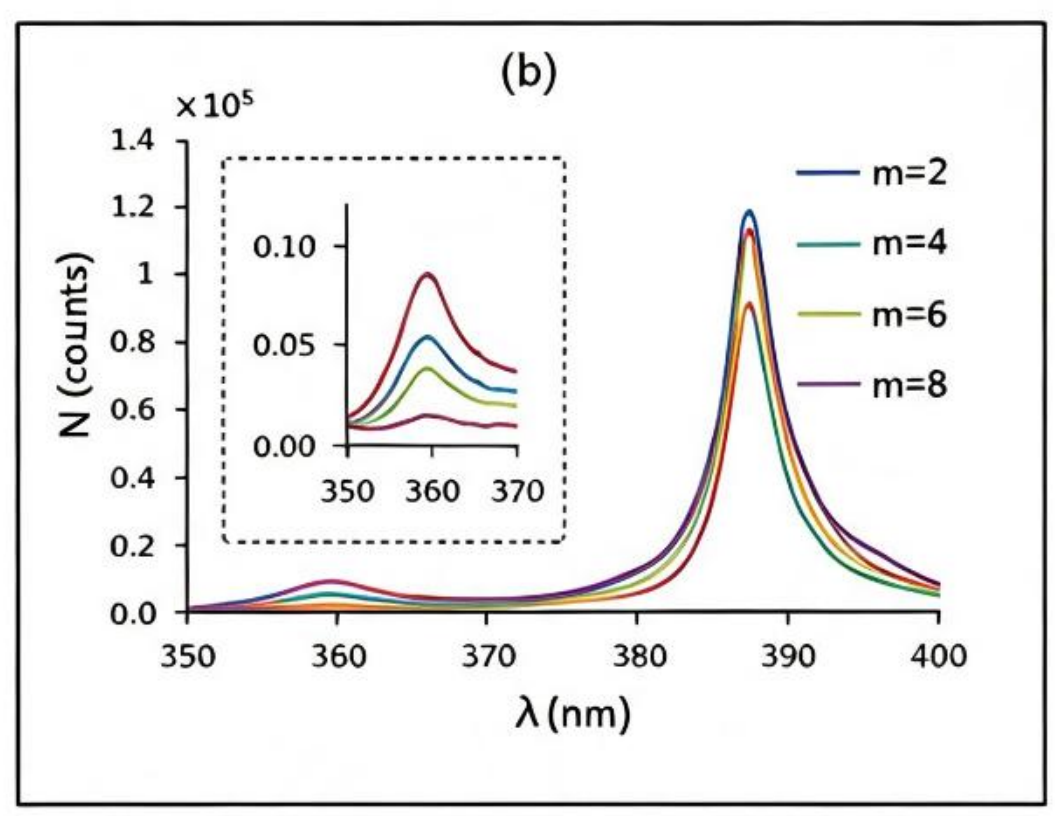

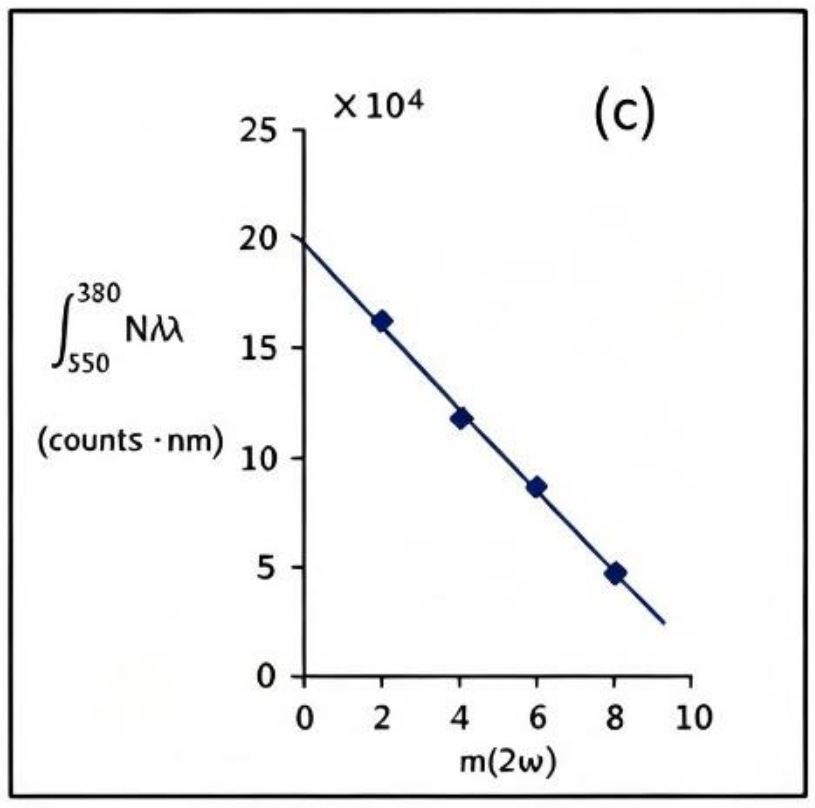

**Fig. 4.** (a) Typical SC spectrum excited at 387.5 nm, spanning from below 300 nm to above 450 nm. (b) Near-threshold SC spectra for topological charges $m_{2\omega}$ = 2, 4, 6, and 8 over 350–400 nm; the anti-Stokes intensity decreases with increasing $m_{2\omega}$. (c) Integrated intensity over 350–380 nm plotted against $m_{2\omega}$, showing a linear decay that confirms the OAM-induced reduction in effective peak intensity as the dominant suppression mechanism.

## 3.4 High-NA inscription and chiral structures

Figure 5 presents side-view transmission optical micrographs of structures inscribed in PMP using an NA ~ 0.4 objective, with linearly polarized helical beams carrying $m_{2\omega} = +2$ at pulse energies from 0.15 to 0.4 μJ and focus depth ~400 μm. At this NA, the ring-shaped LG profile is well resolved ahead of the focal plane, and a visually apparent twist emerges at the focus. This twist reflects the azimuthal asymmetry in energy deposition arising from the transverse Poynting vector of the helical beam: as the wavefront converges, the transverse energy flow redistributes the nonlinear excitation around the focal region, resulting in a rotated or sheared intensity pattern. The full convergence angle for this mode is $\theta_{m=2} \approx 4.1°$, indicating that the geometric phase gradient, while modest, is sufficient to break the reflection symmetry of the inscription profile.

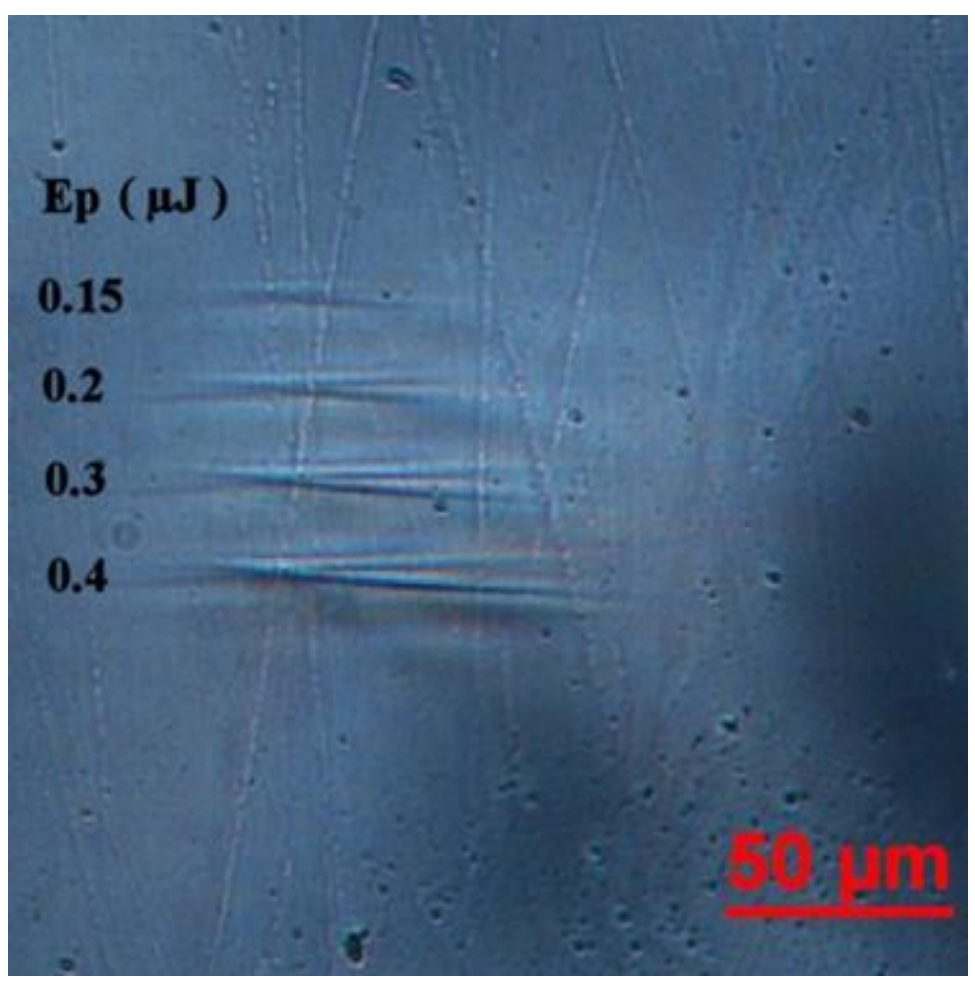

**Fig. 5.** Side-view micrographs of PMP inscriptions (NA = 0.4, $m_{2\omega} = +2$) at 0.15–0.4 μJ and ~400 μm depth. Ring-shaped LG profile ahead of focus; twist at focus reveals azimuthal energy flow. Convergence angle $\theta_{m=2} \approx 4.1°$ confirms symmetry breaking.

To further examine the role of OAM in the high-NA inscription regime, inscriptions were performed with linear and circular polarizations for $m_{2\omega} = \pm 2$ at 0.2 μJ and 0.5 mm/s, as shown in Figs. 7(a) and 7(b). For both helicities, the ring-mode structure near the focus is clearly resolved, yet the polarization state exhibits negligible influence on the overall morphology. This insensitivity confirms that, even at high NA, the non-equilibrium energy flux is governed by the OAM-induced spatial phase gradient. The Gaussian beam, shown for comparison, exhibits stronger coupling due to its higher on-axis intensity, consistent with the annular intensity redistribution that reduces the effective peak intensity for vortex beams. Taken together, these observations support a picture in which the helical phase topology, rather than the polarization state, dictates the nonlinear energy redistribution under high-NA focusing.

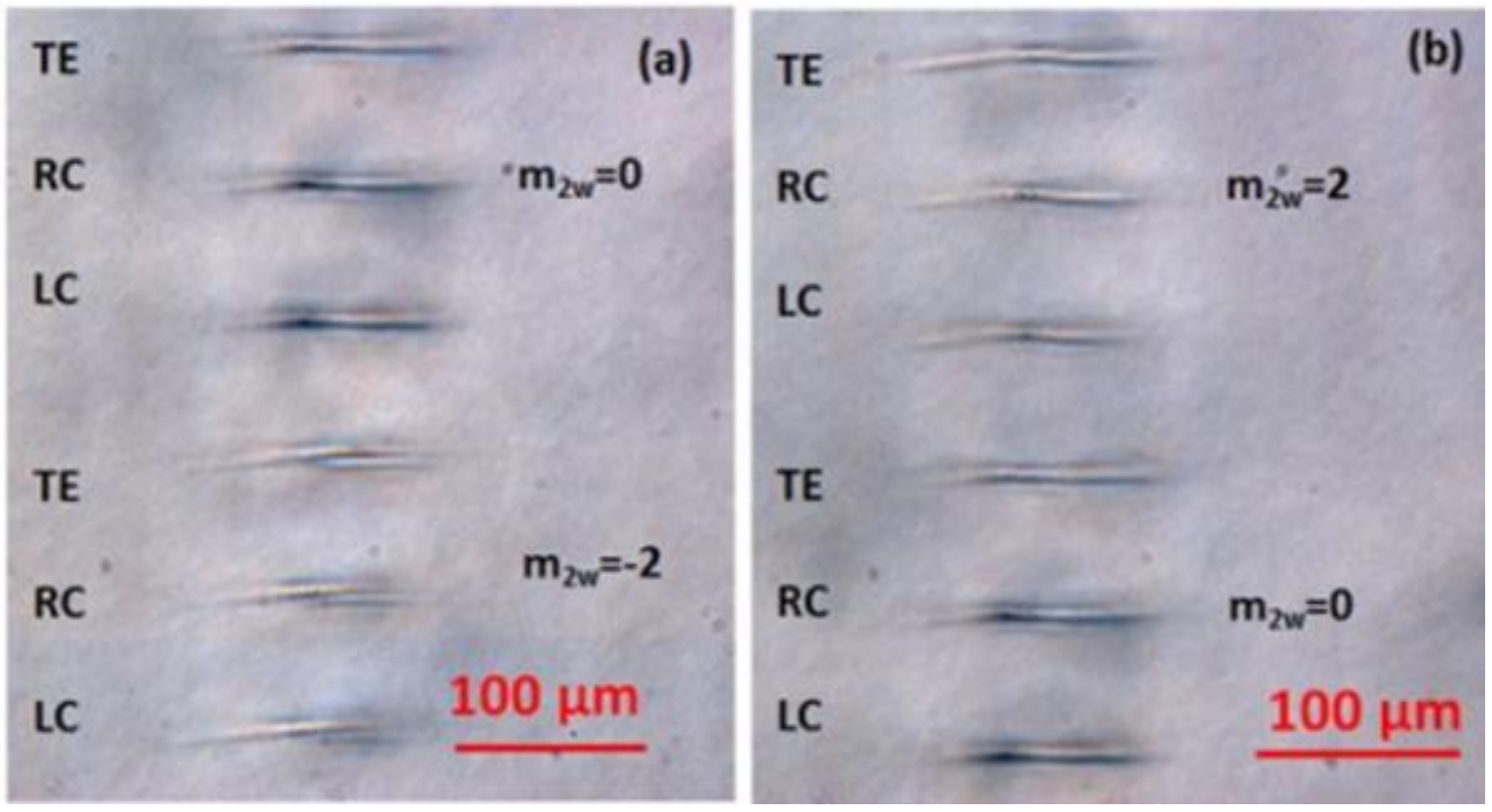

**Fig. 6.** Inscription at 387.5 nm with (a) $m_{2\omega} = -2$ and (b) $m_{2\omega} = +2$, under linear and circular polarizations, compared with a polarized Gaussian beam. Pulse energy was 0.2 µJ, below the self-focusing threshold, using transverse writing at a feed speed of 0.5 mm/s.

Figure 7 illustrates the filamentation structures formed with an OAM beam ($m_{2\omega} = -6$) at NA = 0.4, with a pulse energy of Ep = 0.21 ± 0.005 µJ. Under these conditions, multiple filaments are observed near the focal plane, as shown in Fig. 7(a), with two dominant filaments exhibiting a specific azimuthal orientation at the focus, presented in Fig. 7(c). As the imaging plane shifts deeper into the sample, these two filaments undergo a clockwise rotation, seen in Fig. 7(b). Notably, when the depth is increased further to approximately 30 µm beyond the focus, the rotation reverses to an anti-clockwise direction, as illustrated in Fig. 7(d).

This depth-dependent reversal demonstrates that the azimuthal dynamics of the filaments are not determined by the sign of the OAM alone. Instead, the evolution of the filament pattern along propagation suggests that the local balance among competing processes, such as self-focusing, plasma defocusing, and geometric phase accumulation, varies with longitudinal position [42]. Such variation can effectively modify the transverse energy flow in different regions of the focal volume. The reproducibility of these patterns further indicated that the dissipative dynamics governing filament rotation are deterministic, yet the resulting rotation sense is critically influenced by propagation effects within the absorbing medium.

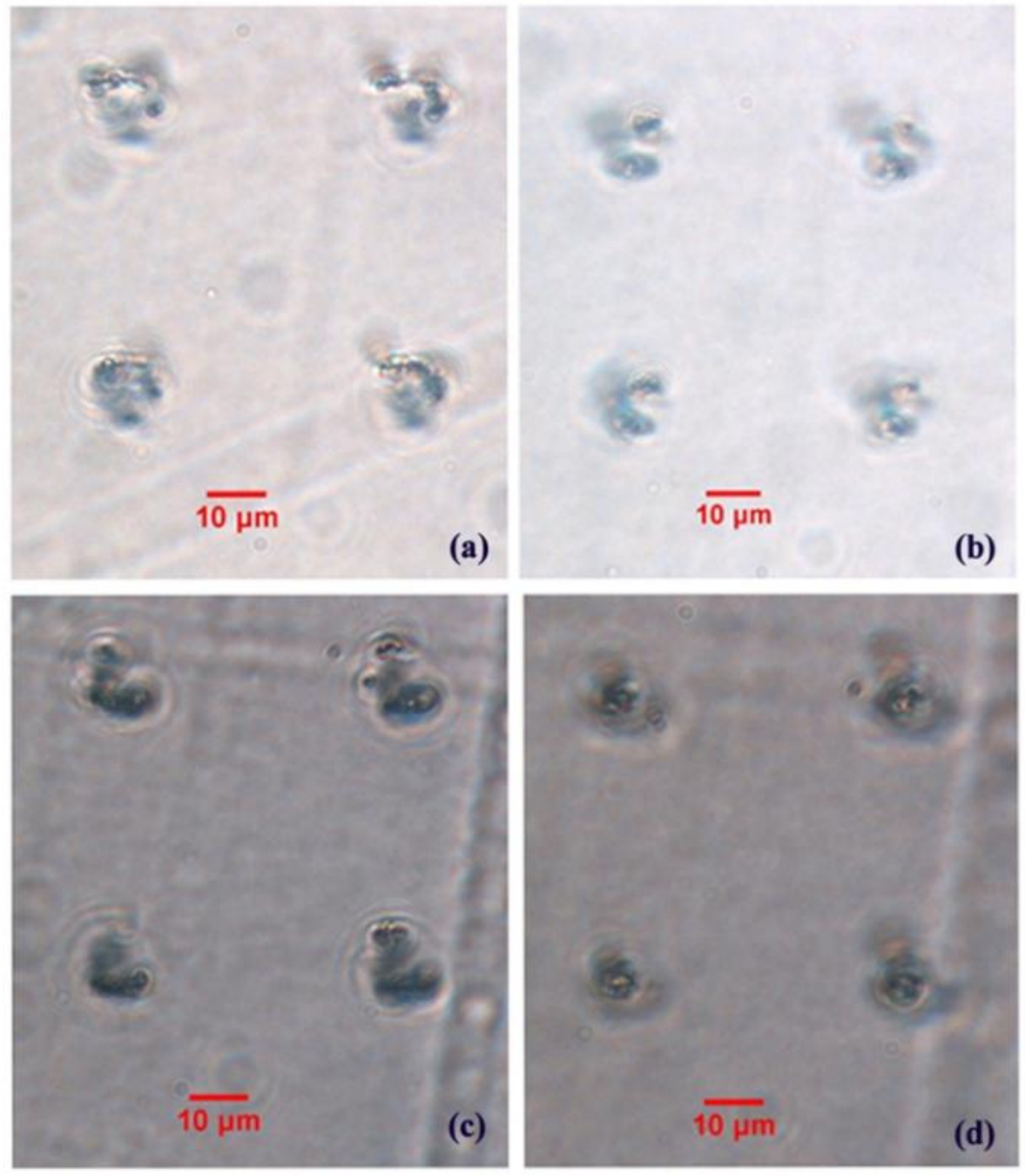

**Fig. 7.** (a) Optical cross-sections of filamentation along the optic axis in PMP for $m_{2\omega} = -6$ at NA = 0.4 (Ep = 0.21 ± 0.005 μJ, 0.2 s). (b) Clockwise rotation of the two dominant filaments with increasing depth. (c) Reference image at the focal plane for $m_{2\omega} = -6$. (d) At ~30 μm deeper focus, rotation reverses to anti-clockwise, showing depth-dependent azimuthal dynamics beyond OAM sign alone.

Increasing the NA further to 0.7 enables waveguide formation just below the surface, where the non-equilibrium energy flux is confined to sub-micrometer dimensions. Figure 8(a) presents transmission images of modification cross-sections inscribed by an OAM beam ($m_{2\omega} = 10$) at pulse energies from 0.3 to 0.5 μJ with an exposure time of 50 ms. At 0.3 μJ, a well-defined light-guiding core of 1.5–2 μm diameter is visible, indicating localized refractive index modification and stable waveguide formation. At 0.5 μJ (Fig. 8(b)), the structure becomes more complex, with multiple filament-guiding regions surrounded by concentric rings. These rings arise from diffraction of the back-illuminating white-light source by the complex refractive index landscape surrounding the modified core. The emergence of multiple guiding regions at higher energy is consistent with the collapse of a high-order OAM beam into multiple filaments, each serving as a localized channel for the non-equilibrium energy flux [43].

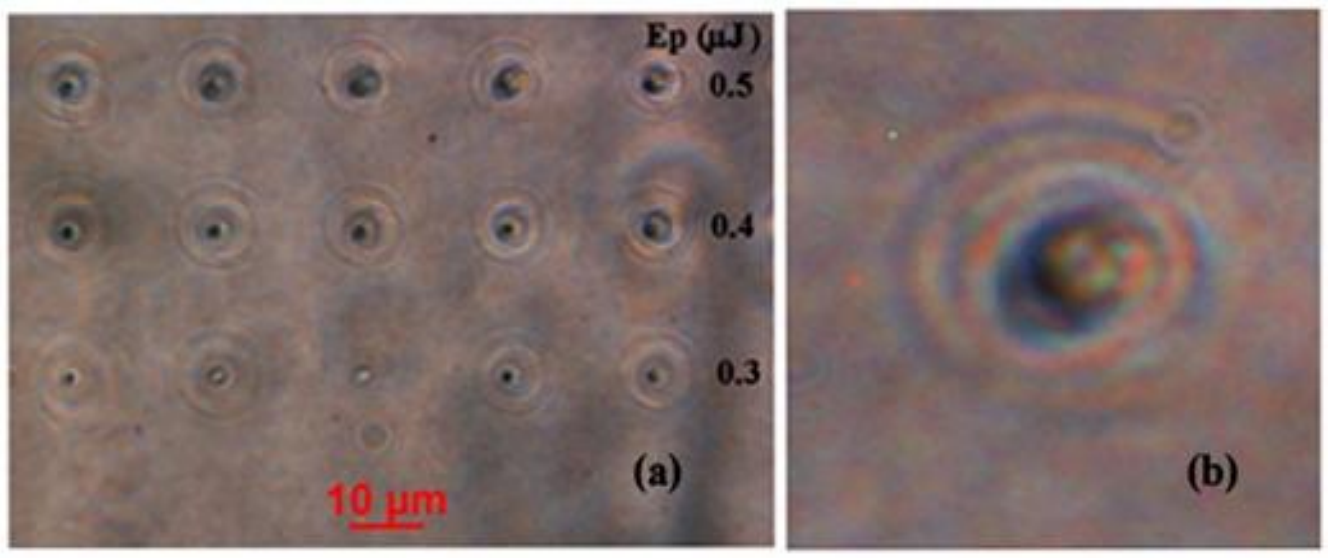

**Fig. 8.** (a) Transverse modification cross-sections in PMP inscribed by OAM beam ($m_{2\omega}$ = 10, NA = 0.7) at 0.3–0.5 μJ, 50 ms. (b) Enlarged view at 0.5 μJ showing multiple guiding regions; surrounding rings arise from white-light diffraction.

Longitudinal waveguides inscribed at NA = 0.7 with $m_{2\omega} = \pm 10$, Ep = 0.5 μJ, and scan speed 0.5 mm/s reveal a striking asymmetry dependent on the sign of the topological charge, as shown in Fig. 9. Structures inscribed with $m_{2\omega} = +10$, presented in Fig. 9(b), are remarkably round and uniform. In stark contrast, those inscribed with $m_{2\omega} = -10$, shown in Fig. 9(c), exhibit a clear twisted, clockwise spiral morphology, a feature absent in the positive OAM case. This helicity-dependent asymmetry is the most prominent signature of the non-equilibrium energy flux in the high-NA, high-energy regime. Unlike in azo-polymer systems [44] or optical tweezers [45], where OAM transfer produces reversible torques and the spiral direction reverses with helicity, the present asymmetry does not reverse sign—suggesting that the mechanism is a non-purely conservative angular momentum transfer but rather a dissipative, non-equilibrium process.

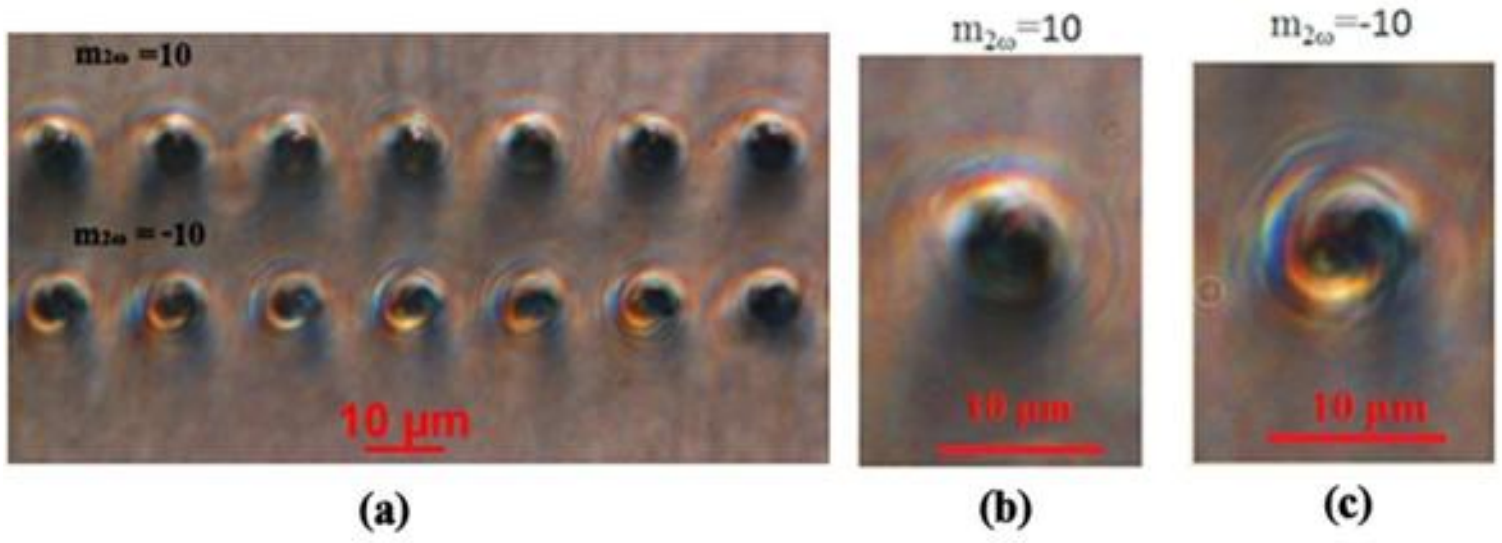

**Fig. 9.** (a) Optical image (transmission) of micro-structures inscribed longitudinally along the optic axis for $m_{2\omega}$=±10, scan speed = 0.5 mm/s. (b) Expanded view of the structure for $m_{2\omega}$=+10. (c) Expanded view of the structure for $m_{2\omega}$=−10, showing a distinct clockwise spiral.

## 3.5 From Low-NA Filamentation to High-NA Helicity-Dependent Inscription: A Unified Tensorial Framework

For analytical tractability, an isotropic material model is adopted as a computational approximation to enable study of the underlying nonlinear dynamics. The framework established in this section predicts that a helical beam with 1 μJ pulse energy and OAM number m = 1 undergoes Kerr-mediated symmetry breaking, redirecting the Poynting vector asymmetrically to generate localized hot spots with power densities of $10^{14}$–$10^{15}$ W/cm², as shown in Fig. 10. Mapping the simulation onto a 12-μm focal spot yields a hot spot diameter of ~0.43μm and area of ~200 nm²; even with only 30% – 50% of the pulse energy concentrated into the hot spot, the power density remains within this range, exceeding most optical material damage thresholds and implying strong plasma formation.

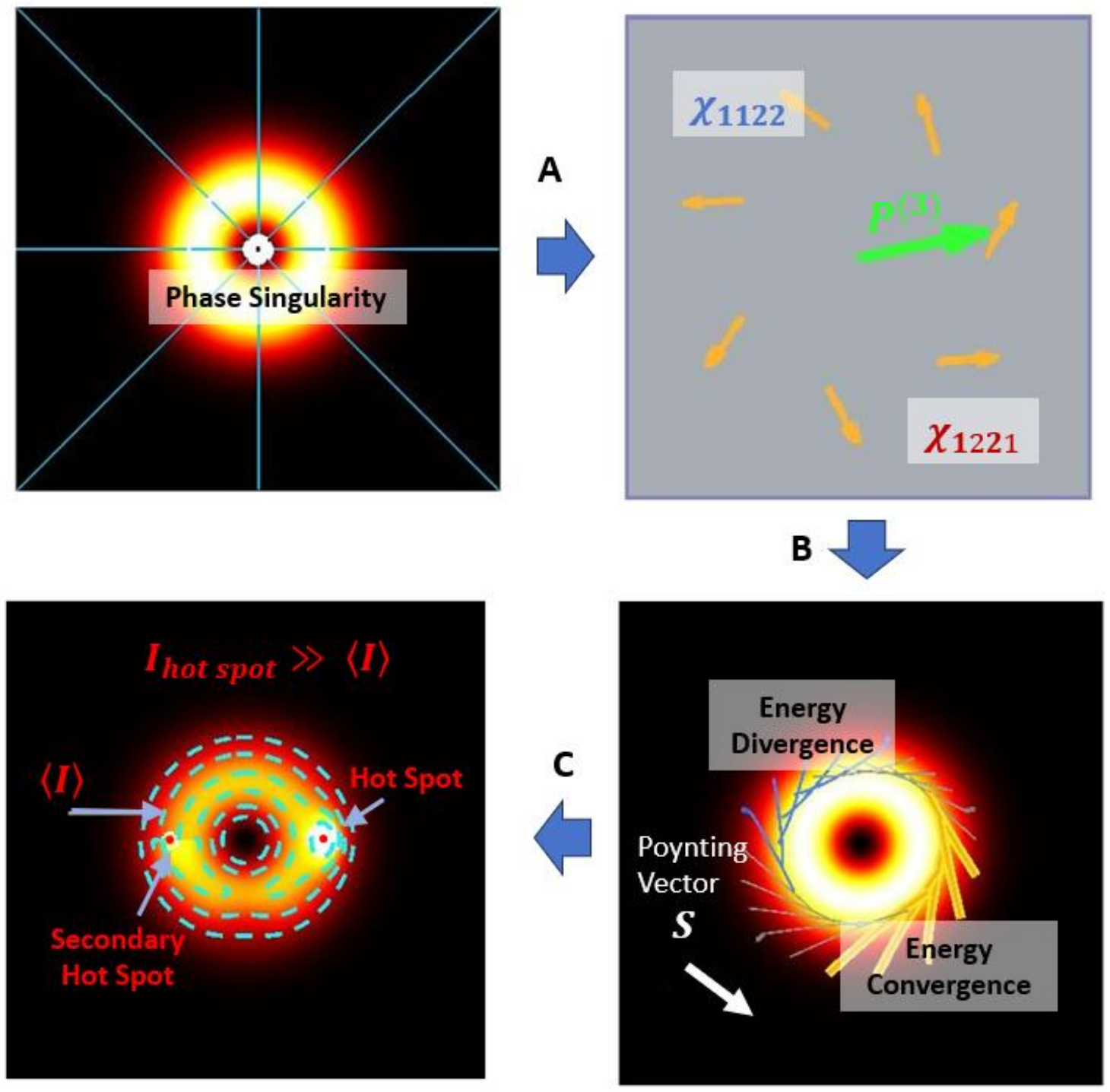

**Fig. 10.** Simulation of localized hot spots (~$10^{14}$–$10^{15}$ W/cm², ~200 nm²) via OAM-induced symmetry breaking, illustrating (a) Kerr nonlinear coupling, (b) transverse nonlinear current, and (c) symmetry breaking and reshaping dynamics.

The total nonlinear polarization induced by the vortex beam consists of two independent tensor components: $\chi_{1122}$, which governs the scalar Kerr response and remains insensitive to the optical phase topology, and $\chi_{1221}$, which mediates the cross-polarized coupling that converts the helical phase gradient into a transverse nonlinear current, breaking cylindrical symmetry and redirecting energy away from symmetric regions.

To interpret these observations, we establish an analytical framework based on the tensorial nonlinear polarization in isotropic media. The energy flow associated with the nonlinear polarization can be decomposed as indicating Equ. (3):

$$S_{\mathrm{NL}} = S_{\mathrm{Kerr}} + S_{\chi 1221} \quad (3)$$

where $S_{\mathrm{Kerr}}$ originates from $\chi_{1122}$ and remains insensitive to the optical phase topology, while $S_{\chi_{1221}}$ arises from $\chi_{1221}$ and carries the helicity-dependent azimuthal asymmetry.

The effective energy flow that determines the final morphology can be written as Equ. (4):

$$S_{eff} = S_0 + \alpha(m) J_{OAM} + \beta(z) P_{prop}$$

with:

- $S_0$: the baseline symmetric energy flow from the Kerr response ( $\chi_{1122}$);

- $J_{OAM} = m\hat{e}_\phi$: the OAM-driven azimuthal flow from $\chi_{1121}$, $\hat{e}_\phi$ azimuthal unit vector (tangential direction around the beam axis).
- $\alpha(m)$: the helicity coupling coefficient (fixed focal conditions);
- $P_{prop}$: propagation-induced azimuthal contribution from self-focusing, plasma defocusing, and geometric phase accumulation;
- $\beta(z)$: depth-dependent weighting factor.

Under fixed focal conditions, $\beta(z) \ll \alpha(m)$, so $S_{eff} \approx S_0 + \alpha(m)J_{OAM}$, yielding helicity-dependent morphology. Along the optic axis, $\beta(z)$ grows with depth; when $|\beta(z)| > |\alpha(m)|$, the propagation term dominates and the rotation sense becomes depth-dependent, independent of the OAM sign.

## 4. Conclusions

This work has established a unified framework connecting the tensorial nature of the third-order nonlinear susceptibility in isotropic media to the OAM-dependent nonlinear response of bulk polymers, demonstrating that the same $\chi^{(3)}$ tensor components govern phenomena ranging from perturbative nonlinear coupling to dissipative material modification across different focusing regimes. The theoretical foundation revealed that the nonlinear polarization depends not merely on scalar intensity but on the full vectorial field topology through two independent tensor components, predicting that the helical phase gradient drives asymmetric Poynting vector flow that generates a transverse nonlinear current and produces localized hot spots with power densities of $10^{14}$–$10^{15}$ W/cm². Experimentally, low-NA filamentation studies demonstrated that OAM and spin angular momentum do not couple measurably, with enhanced coupling under linear polarization attributed to the activation of two tensor components versus one under circular polarization, while supercontinuum spectroscopy revealed a linear decrease in anti-Stokes intensity with increasing topological charge, confirming that the transverse current reduces effective peak intensity and attenuates self-phase modulation. The transition to high-NA inscription revealed the most profound manifestation of this framework, where residual aberrations break cylindrical symmetry and allow the transverse current to couple the helical phase gradient to the local field orientation, producing helicity-dependent morphological asymmetry characterized by twisted spiral structures for $m_{2\omega}=-10$ versus round symmetric profiles for $m_{2\omega}=+10$. The dissipative nature of this process was confirmed by the absence of helicity reversal, attributed to plasma formation breaking time-reversal symmetry. This continuum across regimes, from polarization-dependent coupling without permanent asymmetry at low NA through spectral attenuation without morphological threshold in supercontinuum spectroscopy to dissipative amplification producing permanent chiral structures at high NA, establishes OAM as a controllable parameter for precision polymer structuring and provides a predictive framework for engineering nonlinear light–matter interactions, with direct implications for chiral photonic device fabrication and ultrafast laser material processing.

### Acknowledgements

This study received financial support from the Talent Introduction Research Funds of CQWU (Grant No. R2018SDQ13). The authors gratefully acknowledge Dr. Stuart Edwardson and Prof.

Geoff Dearden at the University of Liverpool for their valuable contributions, and also thank the School of Engineering for providing equipment, instruments, facilities, as well as technical advice, comments, and academic supervision. Sincere thanks go to Prof. Li Lin at the University of Manchester, and to Prof. Zhenghe Zhu and Prof. Guixiong Liu at the South China University of Technology, for their insightful discussions, advice, and guidance.